\documentclass[11pt,twoside]{article}

\usepackage{asp2006}
\usepackage{epsf}
\usepackage{psfig}
\usepackage{lscape}
\usepackage[dvips]{graphicx}

\begin{document}
\title{The galactic stellar nucleation by globular cluster interactions
in dwarf galaxies}   
\author{Shigeki Inoue \& Masafumi Noguchi}   
\affil{Astronomical Institute of Tohoku University, Japan}    

\begin{abstract} 
Dinamical Friction Problem is a long-standing dilemma about globular
 clusters(hereafter,GCs) belonging to dwarf galaxies. The GCs are
 strongly affected by dynamical friction in dwarf galaxies, and presumed
 to fall into the  galactic center. But GCs do exist in dwarf
 galaxies. Recentry, a new solution was proposed. If dwarf galaxies have
 a cored dark matter halo, in which case the effect of dynamical
 friction will be weaken considerably, and GCs are able to survive
 beyond the age of the universe.

In this study, we discussed why does a constant density cored halo cease
 dynamical friction, by means of N-body simulations.
\end{abstract}

\section*{Introduction}
Even in dwarf galaxies belonging to another cluster of galaxies, many
GCs have been observed (Miller et al.1998). But, why these GCs can
survive from strong dynamical friction is a long-standing puzzle. Read
et al.2006(hereafter,R06), by N-body simulations, discovered that
dynamical friction is weekened significantly if dwarf galaxies have a
constant density core in its dark matter halo. They argue that the halo
particles in constant density core become co-rotating with a GC, and
thus the halo particles construct new equilibrium state including the
GC. This equilibrium cease dynamical friction, thus the GC can survive
beyond the age of the universe. 

But, the new equilibrium is very vulnerable to perturbation. If once
the GC orbit is perturbed and the orbital plane is inclined, this
equilibrium will be broken. And dynamical friction force on the GC is
rejuvenated by perturbation in the orbital inclination. This fragile
nature of the new equiliburium is also indicated in R06. In the paper of
R06, they studied only single GC cases. But, of course, real dwarf
galaxies don't necessarily have only one GC, but severals or more. We
expect that the orbits of GCs will be perturbed by gravity of other
GCs. Then we anticipate that some GCs will fall into the galactic center
and form a stellar nucleus.

\section*{Results \& Conclusions}

\begin{figure}[tbp]
 \begin{minipage}{0.5\hsize}
  \begin{center}
   \includegraphics[width=50mm]{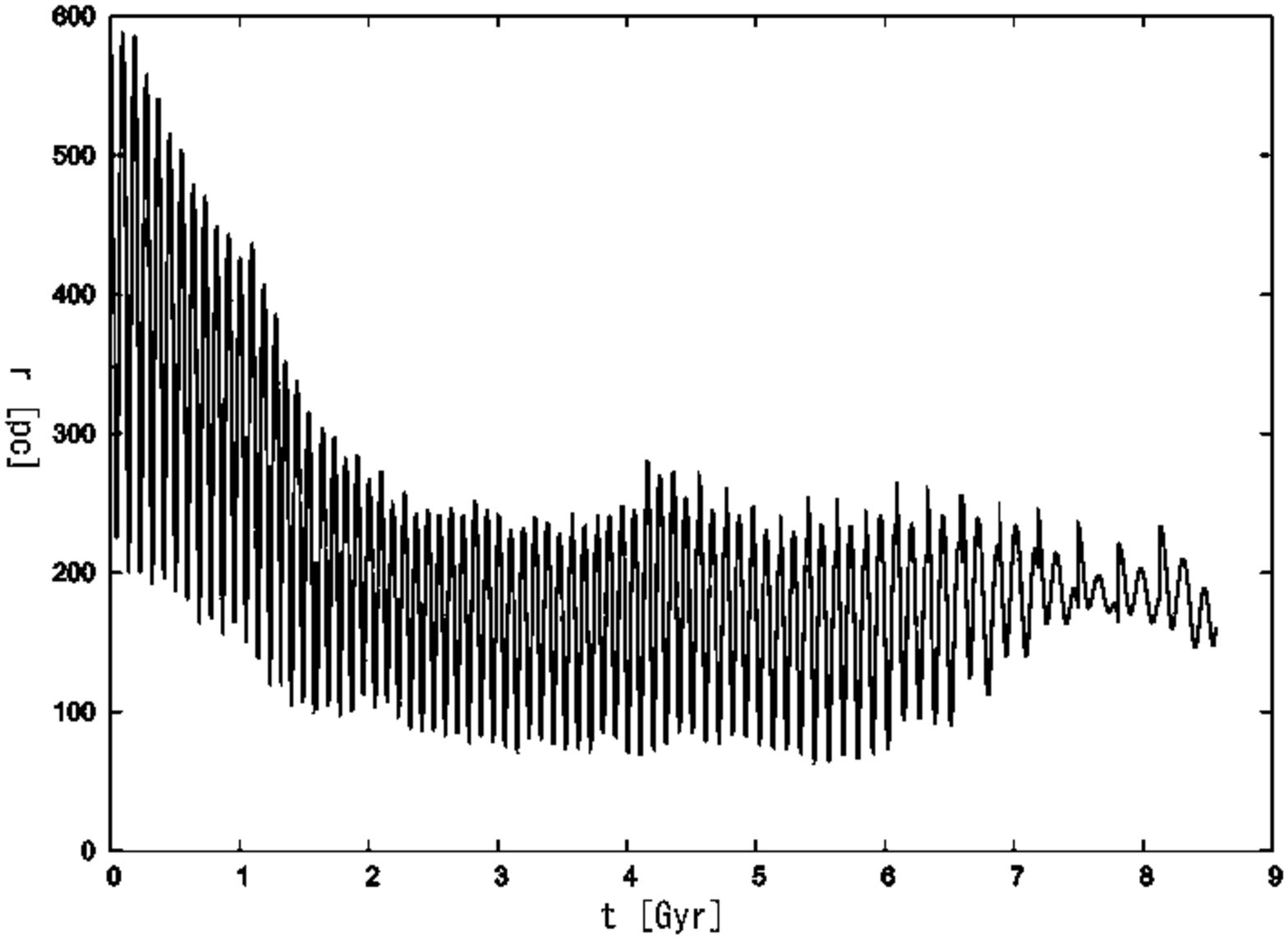}
  \end{center}
  \caption{This is the same simulation as R06.}
  \label{fig:1}
 \end{minipage}
 \begin{minipage}{0.5\hsize}
  \begin{center}
   \includegraphics[width=50mm]{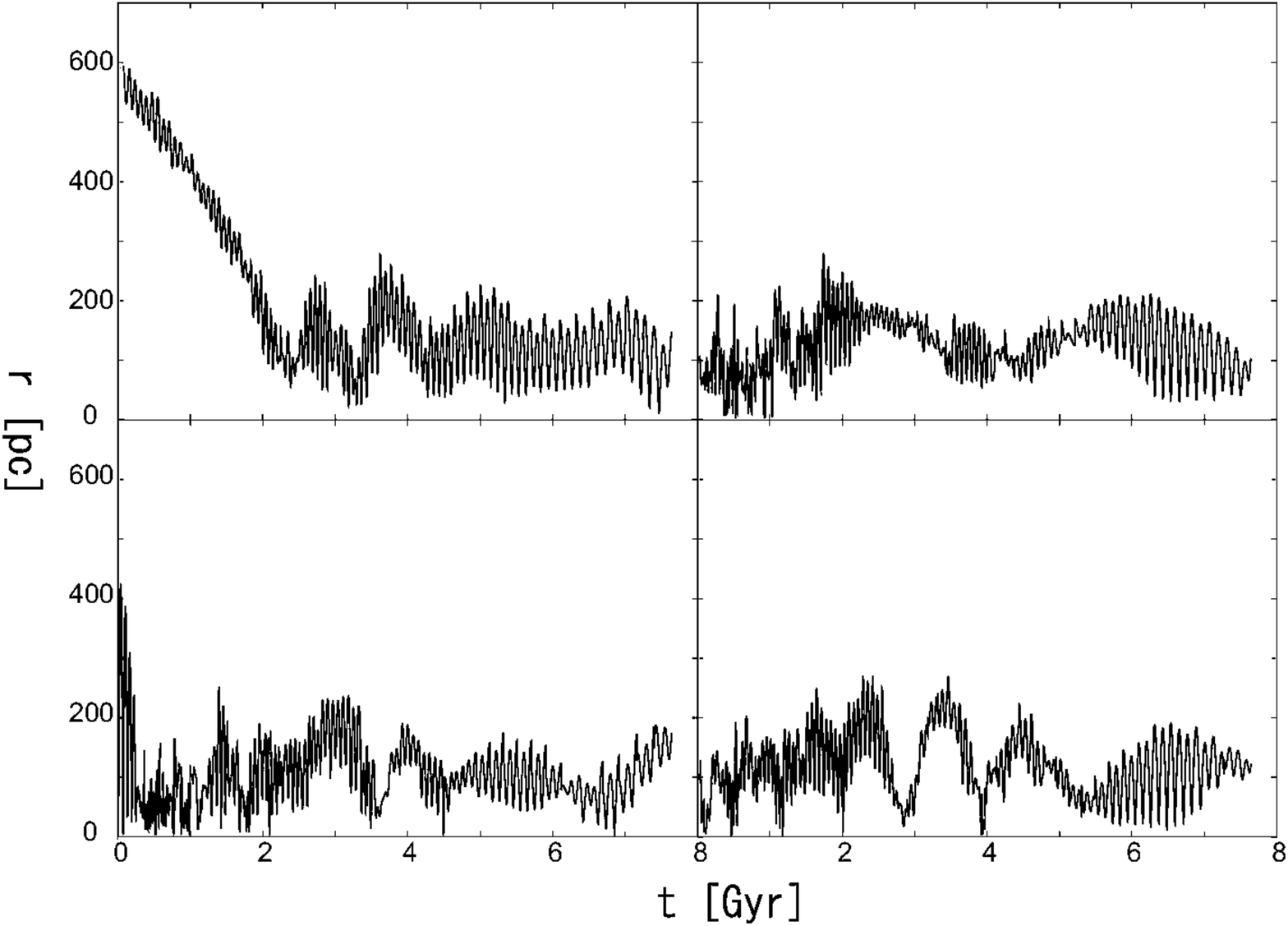}
  \end{center}
  \caption{These are typical results of the 5 GCs case.}
  \label{fig:2}
 \end{minipage}
\end{figure}

First, we carried out the same simulation as R06 in a constant density
cored halo(Fig.\ref{fig:1}). We got the same result as R06. When the GC
entererd the central core region(from 200-300pc), the orbital shrinkage
stopped.

\begin{figure}[tbp]
 \begin{center}
  \includegraphics[width=80mm]{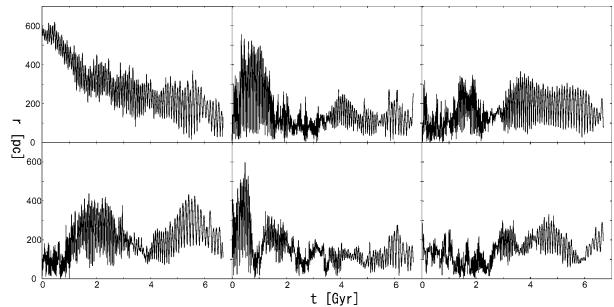}
 \end{center}
 \caption{These are typical results of the 30 GCs case.}
 \label{fig:3}
\end{figure}

Next, we examined 5GCs case and 30GCs case.(Fig.\ref{fig:2} and
Fig.\ref{fig:3}). In comparison with the sigle GC case, clearly GCs are
perturbed by gravitational interaction. Regardless of the nunmber of
GCs, no GC fall into the center of galaxy. All GCs continue orbital
motions in the constant density core region. This result is NOT
consistent with our expectation or the statement of R06.  

In this study, we confirmed that constant a density core ceases
dynamical friction on GCs. This is consistent with the result of
R06. But, even under the perturbation from other GCs, dynamical friction
does NOT affect. This result is inconsistent with the insight of R06. We
consider that this inconsistency is due to the misunderstanding about
the mechanism for suppression of dynamical friction in R06(co-rotating
equilibrium). A constant density core must have another reason of
ceasing dynamical friction. We make a full report on this study in Inoue
\& Noguchi 2008(in preparation).     



\acknowledgements 
The numerical simulations reported here were carried out on GRAPE
systems kindly made available by the Center for Computational
Astrophysics at the National Astronomical Observatory of Japan.


\end{document}